\documentclass[pra,showpacs,showkeys,amsfonts,amsmath,twocolumn]{revtex4}
\usepackage{bm}
\usepackage{graphicx}
\RequirePackage{mathptm}
\newtheorem{thm}{Theorem}

\newtheorem{defn}{Definition}

\newcommand{\si}[1]{\sigma_{#1}}

\newcommand{\ip}[2]{\langle \,{#1},\,{#2}\,\rangle}

\newcommand{\ro}{\varrho}
\newcommand{\la}[1]{\boldsymbol{\lambda}_{#1}}

\newcommand{\al}{\alpha}

\newcommand{\om}{\omega}

\newcommand{\re}{\mathrm{Re}\,}
\newcommand{\I}{\boldsymbol{1}}
\newcommand{\conj}[1]{\overline{#1}}

\newcommand{\cH}{{\mathcal H}}
\newcommand{\cA}{{\mathcal A}}
\newcommand{\cB}{{\mathcal B}}
\newcommand{\Atot}{{\mathcal A}_{\mathrm {tot}}}

\newcommand{\C}{\mathbb C}
\newcommand{\R}{\mathbb R}

\newcommand{\eb}[1]{{\mathbf e}_{#1}}

\newcommand{\im}{\mathrm{Im}}
\newcommand{\cz}[1]{\conj{z}_{#1}}
\newcommand{\z}[1]{z_{#1}}
\newcommand{\tr}{\mathrm{tr}\,}
\begin{document}
\title{Entanglement beyond tensor product structure:
algebraic aspects of quantum non - separability}
\author{{\L}ukasz Derkacz, Marek Gw\'{o}\'{z}d\'{z} and  Lech Jak{\'o}bczyk
}
 \affiliation{Institute of Theoretical Physics\\ University of
Wroc{\l}aw\\
Plac Maxa Borna 9, 50-204 Wroc{\l}aw, Poland}
\begin{abstract}
Algebraic approach to quantum non - separability is applied to the
case of two qubits. It is based on the partition of the algebra of
observables into  independent subalgebras and the tensor product
structure of the Hilbert space  is not exploited. Even in this
simple case, such general formulation has some advantages. Using
algebraic formalism we can explicitly show the relativity of the
notion of entanglement to the observables measured in the system and
characterize separable and non - separable pure states. As a
universal measure of non - separability of  pure states we propose
to take so called total correlation. This quantity depends on the
state as well as on  the algebraic partition. Its numerical value is
given by the norm of the corresponding correlation matrix.
\end{abstract}
\maketitle
\section{Introduction}
Standard approach to the entanglement  of states of distinguishable particles is strictly related to the
tensor product structure of the underlying Hilbert space. In the case of two parties, $\cH=\cH_{1}\otimes
\cH_{2}$ and the state of the system, represented by a density matrix $\ro$, is separable if it can be written as
\begin{equation*}
\ro=\sum\limits_{j}p_{j}\ro^{(1)}_{j}\otimes\ro^{(2)}_{j},\quad
p_{j}\geq 0,\quad \sum\limits_{j}p_{j}=1,
\end{equation*}
where $\ro^{(i)},\, i=1,2$ is the state of $i$th part. Otherwise,
the state $\ro$ is non - separable or entangled. This simple and
natural definition does not work in the case of indistinguishable
particles. The more general notion of quantum non - separability is
needed. This problem was clearly formulated in Ref. \cite{BFM},
where the factorization of the corresponding algebra of observables
into subalgebras describing subsystems, is proposed as a basis for
unambiguous discussion of quantum non - separability. The main idea
is that the questions about entanglement or separability (in the
case of indistinguishable as well as distinguishable particles) are
meaningful only when we specify which statistically independent
(i.e. commuting) subalgebras of the total algebra are considered. In
this formulation, entanglement of quantum states is nothing but the
existence of non - vanishing correlations between such chosen
independent observables. This general definition was earlier used in
mathematically rigorous discussion of quantum correlations and
entanglement in relativistic quantum field theory (see e.g.
\cite{SW, CH, HC}). In such approach one can for example show that
the usual vacuum state is maximally entangled with respect to the
observables localized in complementary wedge - shaped regions in
space - time \cite{SW}.
\par
In the present paper we reconsider the simple case of
distinguishable two qubits using this general algebraic perspective.
We find in particular, that the discussion of relativity of the
notion of entanglement to a set of observables (see e.g. \cite{ZLL,
HR}) is natural and straightforward in the algebraic language. In
this setting the theory is defined by the total algebra of
observables $\Atot$ and the set of states, given by linear
functionals $\om$ on $\Atot$  which attribute to the observables
their expectation values. The subsystems are identified by the
specific choice of subalgebras of $\Atot$ and the tensor product
structure of underlying Hilbert space is not used. To study how the
choice of subsystems influences the entanglement of a given state
$\om$, we need a "universal" measure of quantum non - separability,
not linked to the tensor product structure. In the case of pure
states we propose to use the total correlation in the state $\om$
(see definition below) as such a measure. This measure depends on
the state and the couple of subalgebras, and as we show, is given by
the norm of the corresponding correlation matrix. The last quantity
can be computed by finding the maximal singular value of this
matrix. Moreover, this notion generalizes the Wootters concurrence
\cite{W} in the sense that for the "canonical" choice of subalgebras
directly given by the tensor product structure, the total
correlation and concurrence coincide.
\par
Using the total correlation we show on examples that the same state
on $\Atot$ can have completely different entanglement properties,
depending on the choice of subsystems. Thus the abstract algebraic
formulation of the theory of non - separability is  more natural. In
the algebraic language, the general characterization of separability
of pure states is simple and is given in terms of restrictions of
the state to the subsystems. The notion of restriction of the state
to subalgebra is general and replaces the notion of partial trace
of density matrix, strictly connected to the tensor product
structure. We show that separability is equivalent to the purity of
restrictions. This gives  also the algebraic characterization of
quantum non - separability. In particular, maximal entanglement
follows from the maximal non - purity or trace property of
restrictions.  We finish our study by showing the result previously
established by many authors (see e.g. \cite{HR, TBKN}): any pure
state can be separable or entangled to a given degree of non -
separability, for the appropriate choice of subsystems.

\section{Algebraic framework}
We start with the short review of  the algebraic formulation of
quantum theory (see e.g. \cite{Emch}). Let $\Atot$ be the $\ast$ -
algebra of all observables of the quantum system. Since we consider
the system with finite number of levels, $\Atot$ is isomorphic to
the full matrix algebra. The most general state on $\Atot$ is given
by the linear functional
\begin{equation}
\om\,:\,\Atot\to \C,\label{state}
\end{equation}
which is positive i.e. for all $A\in \Atot$
\begin{equation*}
\om(A^{\ast}A)\geq 0,
\end{equation*}
and normalized  i.e.
\begin{equation*}
\om(\I)=1.
\end{equation*}
If $\Psi$ is a normalized vector in the Hilbert space of the system,
then
\begin{equation}
\om_{\Psi}(A)=\ip{\Psi}{A\Psi}\label{vstate}
\end{equation}
gives the state on $\Atot$ which is called \textit{a vector state}.
On the other hand, for any density matrix $\ro$, the formula
\begin{equation}
\om_{\ro}(A)=\tr (\ro A)\label{mstate}
\end{equation}
defines also the state on $\Atot$ which is in general  mixed. By the
GNS construction \cite{Emch, BR}, every state $\om$ on $\Atot$ is a
vector state on the appropriate Hilbert space $\cH_{\om}$. The set
$\mathbb E$ of all states is convex and compact and the extremal
points of $\mathbb E$ are identified with the pure states of the
system. In the algebraic language the pure states are characterized
by the properties of the GNS representation: $\om$ is pure if and
only if the corresponding representation is irreducible \cite{BR}.
Let $\al$ be $\ast$ - automorphism of $\Atot$ i.e. $\al$ is such
isomorphism of $\Atot$ that $\al(A^{\ast})=\al(A)^{\ast},\, A\in
\Atot$. For every state $\om$ such that $\om_{\al}:=\om\circ \al\neq
\om,\, \om_{\al}$ gives a new state on $\Atot$. Moreover, if $\om$
is pure, then $\om_{\al}$ is also pure.
\par
To describe subsystems of the quantum system, we choose  some
subalgebras of the total algebra $\Atot$. In the context of
separability of quantum states, we consider a pair  $(\cA,\; \cB)$
of isomorphic subalgebras of $\Atot$ with the following properties:
\begin{itemize}
\item[$\bullet$] the subalgebras $\cA$ and $\cB$ are statistically
independent, in the sense that  for all $A\in\cA$ and $B\in\cB,\;
[A,B]=0$,
\item[$\bullet$] the subalgebras $\cA$ and $\cB$ generate the total
algebra i.e. $\Atot=\cA\vee\cB$.
\end{itemize}
Any pair of subalgebras satisfying the above conditions will be
called \textit{a Bell pair} of subalgebras of the total algebra
$\Atot$.
\begin{defn}
Let $(\cA,\,\cB)$ be a Bell pair of subalgebras of $\Atot$. The pure
state $\om$ on $\Atot$ is $(\cA,\,\cB)$ - separable if
\begin{equation*}
\om(AB)=\om(A)\om(B),\quad A\in\cA,\; B\in \cB.
\end{equation*}
A mixed state is $(\cA,\, \cB)$ - separable if it can be expressed
as a convex combination of pure $(\cA,\, \cB)$ - separable states.
\end{defn}
The state $\om$ is $(\cA,\, \cB)$ - \textit{correlated} or
\textit{non - separable} if it is not $(\cA,\, \cB)$ - separable. To
indicate how much a given pure state $\om$ differs from the
separable one for a fixed choice of Bell pair of subalgebras, we
introduce the  quantity which we call \textit{total correlation in
the state} $\om$. It is defined as
\begin{equation}
C_{\om}(\cA,\,\cB)=\sup\,|\om(AB)-\om(A)\om(B)|.\label{korel}
\end{equation}
In the formula (\ref{korel}) the supremum is taken over all
normalized elements $A\in\cA$ and $B\in\cB$. It follows that
\begin{equation*}
0\leq C_{\om}(\cA,\, \cB)\leq 1.
\end{equation*}
In the next section we apply the idea of algebraic non - separability to the case of two qubits.
\section{Two qubits}
\subsection{The total algebra}
Consider the four - level quantum system. It is given by the Hilbert
space $\cH=\C^{4}$ with the canonical basis
$\eb{1},\,\eb{2},\,\eb{3}$ and $\eb{4}$. The total algebra $\Atot$
can be considered as generated by matrix unit $\I$ and elements
$\la{1},\ldots,\la{15}$, where
\begin{equation*}
\la{i}=\I\otimes \si{i},\quad  \la{3+i}=\si{i}\otimes \I,\quad
i=1,2,3
\end{equation*}
and $\la{j},\; j=7,\ldots,15$ are given by Kronecker products of the
Pauli matrices $\si{i}$ taken in the lexicographical order. In the following we will write
\begin{equation*}
\Atot=\left[\,\I,\la{1},\ldots,\la{15}\,\right].
\end{equation*}
An arbitrary element $A\in\Atot$  has the  form
\begin{equation}
A=c_{0}\I+\sum\limits_{j=1}^{15}c_{j}\la{j},\quad
c_{0},\,c_{j}\in\C,\label{Atot}
\end{equation}
so every state is defined by formula
\begin{equation}
\om(A)=c_{0}+\sum\limits_{j=1}^{15}c_{j}w_{j},\label{om4}
\end{equation}
where
\begin{equation}
w_{j}=\om(\la{j}),\quad j=1,\ldots,15
\end{equation}
are the real numbers. In a particular case of pure state $\om_{\Psi}$, where
\begin{equation}
\Psi= z_{1}\eb{1}+\cdots +z_{4}\eb{4},\quad |z_{1}|^{2}+\cdots
+|z_{4}|^{2}=1,\label{psi}
\end{equation}
the sequence $\{w_{j}\}$ is given by
\begin{equation}
\begin{split}
&w_{1}=2\,\re\,(\cz{1}\z{2}+\cz{3}\z{4}),\quad
w_{2}=2\,\im\,(\cz{1}\z{2}+\cz{3}\z{4}),\\[2mm]
&w_{3}=|\z{1}|^{2}-|\z{2}|^{2}+|\z{3}|^{2}-|\z{4}|^{2},\quad
w_{4}=2\,\re\,(\cz{1}\z{3}+\cz{2}\z{4}),\\[2mm]
&w_{5}=-2\,\im\,(\cz{1}\z{3}+\cz{2}\z{4}),\quad
w_{6}=|\z{1}|^{2}+|\z{2}|^{2}-|\z{3}|^{2}-|\z{4}|^{2},\\[2mm]
&w_{7}=2\,\re\,(\cz{2}\z{3}+\cz{1}\z{4}),\quad
w_{8}=2\,\im\,(\cz{2}\z{3}-\cz{1}\z{4}),\\[2mm]
&w_{9}=2\,\re\,(\cz{1}\z{3}-\cz{2}\z{4}),\quad
w_{10}=-2\,\im\,(\cz{2}\z{3}+\cz{1}\z{4}),\\[2mm]
&w_{11}=2\,\re\,(\cz{2}\z{3}-\cz{1}\z{4}),\quad
w_{12}=2\,\im\,(\cz{2}\z{4}-\cz{1}\z{3}),\\[2mm]
&w_{13}=2\,\re\,(\cz{1}\z{2}-\cz{3}\z{4}),\quad
w_{14}=2\,\im\,(\cz{3}\z{4}-\cz{1}\z{2})
\end{split}\label{wpsi}
\end{equation}
and
\begin{equation}
w_{15}=|\z{1}|^{2}-|\z{2}|^{2}-|\z{3}|^{2}+|\z{4}|^{2}.\label{wpsi15}
\end{equation}
\subsection{Bell pair of subalgebras and correlation matrix}
In the case of two qubits, it is convenient to take subalgebras $\cA$ and $\cB$ defined in the following
way. Let $A_{1},\, A_{2},\, A_{3}$ and $B_{1},\, B_{2},\, B_{3}$ be the linearly independent hermitian elements
of $\Atot$, which satisfy
\begin{equation}
A_{i}^{2}=B_{i}^{2}=\I,\quad [A_{i},B_{j}]=0,\quad i,j=1,2,3\label{AB1}
\end{equation}
and
\begin{equation}
\{A_{i},A_{j}\}=\{B_{i},B_{j}\}=0,\quad i\neq j,\, i,j,=1,2,3\label{AB2}
\end{equation}
We put
\begin{equation}
\cA=\left[\, \I,\, A_{1},\, A_{2},\, A_{3}\,\right],\quad
\cB=\left[\, \I,\,B_{1},\, B_{2},\, B_{3}\,\right]. \label{AB}
\end{equation}
Consider the elements $A\in \cA,\, B\in \cB$ defined as
\begin{equation}
A=a_{1}A_{1}+a_{2}A_{2}+a_{3}A_{3},\quad
B=b_{1}B_{1}+b_{2}B_{2}+b_{3}B_{3}.\label{defAB}
\end{equation}
where $\vec{a}=(a_{1},a_{2},a_{3}),\; \vec{b}=(b_{1},b_{2},b_{3})$ are the real vectors.
By conditions (\ref{AB1}) and (\ref{AB2})
\begin{equation*}
A^{2}=||\vec{a}||^{2}\,\I,\quad B^{2}=||\vec{b}||^{2}\,\I,
\end{equation*}
so if the vectors $\vec{a}$ and $\vec{b}$ are normalized,  $A^{2}=\I$ and $B^{2}=\I$.
From now on we will always assume that $||\vec{a}||=||\vec{b}||=1$. Let $\om$ be an arbitrary pure state on $\Atot$.
Notice that for $A$ and $B$ defined by (\ref{defAB})
\begin{equation}
\om(AB)-\om(A)\om(B)=\ip{\vec{a}}{Q\vec{b}},
\end{equation}
where \textit{the correlation matrix} $Q=(q_{ij})$ has the matrix elements
\begin{equation*}
q_{ij}=\om(A_{i}B_{j})-\om(A_{i})\om(B_{j}).
\end{equation*}
Thus
\begin{equation}
C_{\om}(\cA,\,\cB)=\sup\limits_{\vec{a},\;\vec{b}}\,|\ip{\vec{a}}{Q\vec{b}}|=||Q||,
\end{equation}
where the supremum is taken over all normalized vectors $\vec{a},\, \vec{b}\in\R^{3}$. Thus in the case
of two qubits, the total correlation in the pure state $\om$ can be computed by finding
the matrix norm of the corresponding correlation matrix $Q$ i.e. the largest singular value of $Q$.
\subsection{Canonical Bell pair and concurrence}
The most natural choice of Bell pair is obtained by considering the
subalgebras
\begin{equation}
\cA_{0}=\left[\, \I,\, \la{1},\, \la{2},\, \la{3}\,\right],\quad
\cB_{0}=\left[\, \I,\, \la{4},\, \la{5},\,
\la{6}\,\right].\label{A0B0}
\end{equation}
All conditions defining a Bell pair are trivially satisfied. Notice
also that $(\cA_{0},\,\cB_{0})$ - correlated states can be
identified with standard entangled states with respect to the
partition $\C^{4}=\C^{2}\otimes\C^{2}$, therefore the Bell pair
given by (\ref{A0B0}) will be called \textit{canonical Bell pair} of
subalgebras. An interesting link between the algebraic theory of non
- separability and standard theory of entanglement is given by the
following result (established in the another context by Verstraete
et al \cite{V}):
\begin{thm}
Let $(\cA_{0},\,\cB_{0})$ be the canonical Bell pair of subalgebras of the total algebra of two - qubit system.
For an arbitrary pure state $\om$
\begin{equation*}
C_{\om}(\cA_{0},\, \cB_{0})=C(\om),
\end{equation*}
where $C(\om)$ is the concurrence of $\om$.
\end{thm}
\textit{Proof.} We show this result in the case of vector states $\om_{\Phi}$ defined by
\begin{equation}
\Phi=\sqrt{\frac{1-d}{2}}\,\eb{2}+\sqrt{\frac{1+d}{2}}\,\eb{3},\quad
d=\sqrt{1-c^{2}}\label{Fi}
\end{equation}
where $c\in [0,1]$. One can check that the Wootters concurrence
\cite{W} of the state $\om_{\Phi}$ is equal to $c$. On the other
hand, one easily verifies that the correlation matrix  of
$\om_{\Phi}$ with respect to the canonical Bell pair is given by
$Q_{0}=\mathrm{diag}\,(c,c,-c)$, so
$$
||Q_{0}||=c.
$$
For the general pure state (\ref{psi}) the proof is
analogous.
\subsection{ Non - canonical Bell pairs and relativity of entanglement}
We are going to show by considering the explicit examples that the
notion of entanglement is highly non - unique. Non - separability of
a state is always relative to the measurement setup, which is fixed
by the specific choice  of observables, forming statistically
independent subalgebras $\cA$ and $\cB$.
\par
We start the discussion of this point considering the states which are obviously separable
with respect to the canonical subalgebras $\cA_{0}$ and $\cB_{0}$. Take the vector states defined by the
basic vectors $\eb{1},\, \eb{2},\, \eb{3}$ and $\eb{4}$, but  consider observables belonging to another
subalgebras of $\Atot$. In this case, let $\cA$ and $\cB$ be defined as follows
\begin{equation}
\begin{split}
\cA&=\bigg[\, \I,\, \frac{1}{\sqrt{2}}(\la{4}+\la{11}),\,
\frac{1}{\sqrt{2}}(\la{10}-\la{12}),\\
&\hspace*{4mm}-\frac{1}{2}(\la{1}+\la{3}-\la{13}+\la{15})\,\bigg],\\[2mm]
\cB&=\bigg[\,\I,\, \frac{1}{\sqrt{2}}(\la{7}+\la{9}),\,
-\frac{1}{\sqrt{2}}(\la{5}+\la{8}),\\
&\hspace*{6mm}\frac{1}{2}(\la{1}-\la{3}-\la{13}-\la{15})\,\bigg].
\end{split}\label{AB}
\end{equation}
Using the relations between the generators $\la{j}$, one can easily check that $(\cA,\, \cB)$ is a Bell pair.
To find the correlation matrices corresponding the the considered states, we can utilize the formulas
(\ref{wpsi}) and (\ref{wpsi15}), and we obtain the result:  the states $\eb{1}$ and $\eb{2}$ have
correlation matrices with all zero elements, but  the  correlation matrices corresponding to $\eb{3}$ and $\eb{4}$ are
given by
\begin{equation*}
Q_{\eb{3}}=\mathrm{diag} \,(1,1,-1),\quad Q_{\eb{4}}=\mathrm{diag} \,(-1,-1,-1),
\end{equation*}
and
\begin{equation*}
||Q_{\eb{3}}||=||Q_{\eb{4}}||=1.
\end{equation*}
Thus the states $\eb{1}$ and $\eb{2}$ are $(\cA,\, \cB)$ -
separable, whereas the states $\eb{3}$ and $\eb{4}$ are maximally
$(\cA,\,\cB)$ - entangled.
\par
Consider now the family of states which are maximally entangled with
respect to canonical subalgebras $\cA_{0}$ and $\cB_{0}$. As it is
known \cite{BJO}, such property have the states
$\om_{a,\varphi,\vartheta}$, defined by vectors
\begin{equation}
\Psi_{a,\varphi,\vartheta}=A\,\eb{1}+B\,e^{i\varphi}\,\eb{2}+
B\,e^{i\vartheta}\eb{3}-A\,e^{i(\varphi+\vartheta)}\,\eb{4},\label{maxent}
\end{equation}
where
$$
A=\frac{a}{\sqrt{2}},\quad B=\sqrt{\frac{1-a^{2}}{2}},
$$
and $a\in [0,1],\: \varphi,\:\vartheta \in [0,2\pi]$. This time we
ask about entanglement properties of $\om_{a,\varphi,\vartheta}$ but
with respect to the experimental setup given by the pair
$(\cA^{\prime},\, \cB^{\prime})$ defined below
\begin{equation}
\begin{split}
\cA^{\prime}&=\bigg[\,\I,\,
-\frac{1}{2}(\la{3}-\la{6}-\la{7}+\la{11}),\\
&\hspace*{4mm}-\la{10},\, -\frac{1}{2}(\la{3}+\la{6}-\la{7}-\la{11})\,\bigg],\\[2mm]
\cB^{\prime}&=\left[\,\I,\,
\frac{1}{\sqrt{2}}(\la{1}-\la{9}),-\frac{1}{\sqrt{2}}(\la{5}-\la{14}),\,
\la{15}\,\right].
\end{split}\label{A1B1}
\end{equation}
Again, using (\ref{wpsi}) and (\ref{wpsi15}), we are able to find
the corresponding correlation matrix. This matrix has elements
\begin{equation*}
\begin{split}
&q_{11}=\sqrt{2}\,a\sqrt{1-a^{2}}\,\left(\cos\varphi-a^{2}\,\cos\vartheta\cos(\varphi+\vartheta)\right),\\[2mm]
&q_{12}=\sqrt{2}\,a\sqrt{1-a^{2}}\,\left(\sin\vartheta+a^{2}\,\sin\varphi\cos (\varphi+\vartheta)\right),\\[2mm]
&q_{13}=-2\,a^{2}\,(1-a^{2})\,\cos (\varphi+\vartheta),\\[2mm]
&q_{21}=\sqrt{2}\,a\sqrt{1-a^{2}}\,\sin\vartheta\,\left(2\,a^{2}\,\cos
\varphi\cos\vartheta-
\cos (\varphi-\vartheta)\right),\\[2mm]
&q_{22}=\sqrt{2}\,a\sqrt{1-a^{2}}\,\cos\varphi\,\left(\cos\,(\varphi-\vartheta)
-2a^{2}\sin\varphi\cos\vartheta\right),\\[2mm]
&q_{23}=4a^{2}\,(1-a^{2})\,\cos\varphi\sin\vartheta,\\[2mm]
&q_{31}=\sqrt{2}\,a\sqrt{1-a^{2}}\,\big(\sin\vartheta\,\sin\,(\varphi-2\vartheta)\\
&\hspace*{6mm}-(a^{2}/2)\,
(\cos\varphi +\cos\,(\varphi-2\vartheta)\big),\\[2mm]
&q_{32}=\sqrt{2}\,a\sqrt{1-a^{2}}\,\left(a^{2}\,\sin\varphi\,\cos\,(\varphi-\vartheta)-\cos\varphi\,
\sin\,(\varphi-\vartheta)\right),\\[2mm]
&q_{33}=-2\,a^{2}\,(1-a^{2})\,\cos\,(\varphi-\vartheta).
\end{split}
\end{equation*}
Thus
\begin{equation}
C_{\om_{a,\varphi,\vartheta}}=||Q||=\sqrt{a^{2}(1-a^{2})\,(2+\cos
2\,\varphi-\cos 2\,\vartheta)},\label{Kora}
\end{equation}
and we see that all  states $\om_{a,\varphi,\vartheta}$ with $a=0$
or $a=1$ and $\varphi,\, \vartheta$ arbitrary, are
$(\cA^{\prime},\cB^{\prime})$ - separable. The same property have
the states with $\varphi=\pi/2,\, \vartheta =0$ and any $a\in
(0,1)$. On the other hand, the state defined by the vector
\begin{equation*}
\Psi_{\frac{1}{\sqrt{2}},\,0,\,\frac{\pi}{2}}=\frac{1}{2}\left(\eb{1}+\eb{2}+i\,\eb{3}-i\,\eb{4}\,\right)
\end{equation*}
 gives the maximal value of the norm (\ref{Kora}), so
it is not only  maximally correlated with respect to the canonical
Bell pair $(\cA_{0},\,\cB_{0})$ but also with respect to the pair
$(\cA^{\prime},\,\cB^{\prime})$. So we see that depending on the
experimental setup, separable states can be maximally entangled and
vice versa: maximally entangled states can be separable.
\section{Algebraic  non - separability of two  qubits}
\subsection{Characterization of pure separable states}
Algebraic approach to quantum non - separability consequently takes
into account the relativity of this notion shown by above
discussion. To avoid ambiguities in deciding if a given state is
separable or entangled, we should always specify which statistically
independent subalgebras of the total algebra of observables are
considered. In this general setting we propose to take the total
correlation $C_{\om}(\cA,\cB)$ in a given state $\om$ as an
universal measure of non - separability of pure states. This
quantity does not depend on the tensor product structure of the
underlying Hilbert space and in the case of two qubits and canonical
pair of subalgebras it coincides with  the Wootters concurrence.
Thus the state $\om$ is $(\cA,\, \cB)$ - separable if
$C_{\om}(\cA,\,\cB)=0$ and it is $(\cA,\, \cB)$ - correlated (or
entangled) if $C_{\om}(\cA,\,\cB)=c>0$. Although the total
correlation can be expressed by the norm of the corresponding
correlation matrix, in general it is not easy to find the states for
which this norm vanishes or is greater then zero. We need another
characterization of separability. In the standard approach based on
the tensor product structure of the Hilbert space such
characterization is given by the notion of the partial trace of the
state. In the algebraic setting, the partial trace can be replaced
by the more general notion of \textit{restriction of a state to the
subalgebra}, and we have the following result:
\begin{thm}\label{sep}
Let $\om$ be the pure state on $\Atot$ and let   $(\cA,\, \cB)$ be a Bell pair of subalgebras of $\Atot$.
$\om$ is $(\cA,\, \cB)$ - separable, if and only if the restrictions  $\om_{\cA}$ and $\om_{\cB}$ of $\om$
to $\cA$ and $\cB$
respectively, are also pure.
\end{thm}
\textit{Proof.} This theorem is a particular case of a general result given by Takesaki \cite{Tak},
but for the readers convenience,
we sketch  the proof. Obviously, if $\om$ is separable and pure, then $\om_{\cA}$ and $\om_{\cB}$ are pure.
Assume now that the restriction
$\om_{\cA}$ is pure. Let $A\in \cA$ and $B\in \cB$. Without the loss of generality, we can consider only such
$B\in \cB$, that
$0\leq B\leq \I$.
Assume also that $0<\om(B)<1$, then for $A\in \cA$ we have
\begin{equation}
\om_{\cA}(A)=\om(B)\,\frac{1}{\om(B)}\,\om(AB)+(1-\om(B))\,\frac{1}{1-\om(B)}\,\om(A(\I-B))\label{convkom}
\end{equation}
Since $B$ commutes with all elements of $\cA$, the functionals
\begin{equation*}
\om_{1}(A)=\frac{1}{\om(B)}\,\om(AB),\quad \om_{2}(A)=\frac{1}{1-\om(B)}\,\om(A(\I-B))
\end{equation*}
are both states of $\cA$. Indeed, $\om_{1}(\I)=\om(\I)=1$ and
\begin{equation*}
\om_{1}(A^{\ast}A)=\frac{1}{\om(B)}\,\om(A^{\ast}AB)=\frac{1}{\om(B)}\,\om((B^{1/2}A)^{\ast}\,B^{1/2}B))\geq 0
\end{equation*}
and similarly for $\om_{2}$. So the equality (\ref{convkom}) means that $\om_{\cA}$ is a convex combination of other
states of $\cA$, but by the assumption $\om_{\cA}$ is pure, hence
\begin{equation*}
\om_{\cA}(A)=\om_{1}(A)=\om_{2}(A)
\end{equation*}
i.e.
\begin{equation*}
\om_{\cA}(A)=\frac{\om(AB)}{\om(B)}
\end{equation*}
and $\om$ is $(\cA,\, \cB)$ - separable. \\[2mm]
In the case of two qubits, the condition of separability given by
the above theorem can be reformulated as follows. Let
\begin{equation*}
\cA=\{\,\I,A_{1},\, A_{2},\, A_{3}\,\},\quad \cB=\{\,\I,\, B_{1},\,B_{2},\, B_{3}\,\}
\end{equation*}
be any Bell pair of subalgebras. The restrictions $\om_{\cA}$ and
$\om_{\cB}$ are equivalent to one qubit states, so the states
$\om_{\cA}$ and $\om_{\cB}$ are pure if the real vectors
$\vec{r}_{\om}$ and $\vec{s}_{\om}$, defined by
\begin{equation}
\vec{r}_{\om}=(\om(A_{1}),\, \om(A_{2}),\,\om( A_{3})),\quad
\vec{s}_{\om}=(\om(B_{1}),\, \om(B_{2}),\, \om(B_{3})), \label{rs}
\end{equation}
satisfy
\begin{equation}
||\vec{r}_{\om}||=1,\quad ||\vec{s}_{\om}||=1. \label{purerestr}
\end{equation}
Using the condition (\ref{purerestr}), we can find all separable
states for a fixed pair of subalgebras $\cA$ and $\cB$. To clarify
this point and give an example,  take the Bell pair
$(\cA^{\prime},\, \cB^{\prime})$ defined by (\ref{A1B1}) and
consider the most general pure state $\om_{\Psi}$. By (\ref{wpsi})
and (\ref{wpsi15}) we have that the values of the functional
$\om_{\Psi}$ on the generators $A_{i}$ of $\cA^{\prime}$ are given
by
\begin{equation}
\begin{split}
&\om_{\Psi}(A_{1})=|z_{2}|^{2}-|z_{3}|^{2}+2\,\re\, \conj{z}_{1}\,z_{4},\\
 &\om_{\Psi}(A_{2})=-2\,\im\,(\conj{z}_{2}z_{3}+\conj{z}_{1}z_{4}),\\
 &\om_{\Psi}(A_{3})=|z_{4}|^{2}-|z_{1}|^{2}+2\,\re\,
 \conj{z}_{2}z_{3}.
\end{split}\label{omA}
\end{equation}
Analogously, for the generators $B_{i}$ of $\cB^{\prime}$
\begin{equation}
\begin{split}
&\om_{\Psi}(B_{1})=\sqrt{2}\,\re\left(\conj{z}_{1}(z_{2}-z_{3})+(\conj{z}_{2}+\conj{z}_{3})z_{4}\right),\\
&\om_{\Psi}(B_{2})=\sqrt{2}\,\im \left(\conj{z}_{1}(z_{3}-z_{2})+(\conj{z}_{2}+\conj{z}_{3})z_{4}\right),\\
&\om_{\Psi}(B_{3})=|z_{1}|^{2}-|z_{2}|^{2}-|z_{3}|^{2}+|z_{4}|^{2}.
\end{split}\label{omB}
\end{equation}
 It is not easy task to obtain a general solution of the conditions
(\ref{purerestr}), so we look for "basic" separable states $\om$,
for which the vectors $\vec{r}_{\om}$ and $\vec{s}_{\om}$ have only
one non - zero component which is equal to $+1$ or $-1$. Analyzing
the  expressions (\ref{omA}) and (\ref{omB}), we find the simple
solutions of conditions (\ref{purerestr}) which correspond to the
canonical basis of $\C^{4}$. In particular we obtain that:
\begin{itemize}
\item[$\bullet$]
$\vec{r}_{\om}=(0,0,-1),\, \vec{s}_{\om}=(0,0,1)$ give the  state
$\eb{1}$,
\item[$\bullet$]
$\vec{r}_{\om}=(1,0,0),\, \vec{s}_{\om}=(0,0,-1)$ give the  state
$\eb{2}$,
\item[$\bullet$]
$\vec{r}_{\om}=(-1,0,0),\, \vec{s}_{\om}=(0,0,-1)$ give the  state
$\eb{3}$,
\item[$\bullet$]
$\vec{r}_{\om}=(0,0,1),\, \vec{s}_{\om}=(0,0,1)$ give the  state
$\eb{4}$.
\end{itemize}
Hence the vectors $\eb{1},\, \eb{2},\, \eb{3}$ and $\eb{4}$ are not
only $(\cA_{0},\, \cB_{0})$ - separable, but they are also
$(\cA^{\prime},\, \cB^{\prime})$ - separable.\\ Other solution of
(\ref{purerestr}) allow to construct interesting examples of "non -
standard" separable vector states. In particular we find that:
\begin{itemize}
\item[$\bullet$]
$\vec{r}_{\om}=(-1,0,0),\, \vec{s}_{\om}=(0,0,1)$ give the state
\begin{equation*}
\Psi_{-}=\frac{1}{\sqrt{2}}\left(\eb{1}-\eb{2}\right),
\end{equation*}
\item[$\bullet$]
$\vec{r}_{\om}=(1,0,0),\, \vec{s}_{\om}=(0,0,1)$ give the state
\begin{equation*}
\Psi_{+}=\frac{1}{\sqrt{2}}\left(\eb{1}+\eb{2}\right),
\end{equation*}
\item[$\bullet$]
$\vec{r}_{\om}=(0,0,-1),\, \vec{s}_{\om}=(0,0,-1)$ give the state
\begin{equation*}
\Phi_{-}=\frac{1}{\sqrt{2}}\left(\eb{2}-\eb{3}\right),
\end{equation*}
\item[$\bullet$]
$\vec{r}_{\om}=(0,0,1),\,\vec{s}_{\om}=(0,0,-1)$ give the state
\begin{equation*}
\Phi_{+}=\frac{1}{\sqrt{2}}\left(\eb{2}+\eb{3}\right).
\end{equation*}
\end{itemize}
Notice that  $\Psi_{-},\, \Psi_{+},\, \Phi_{-}$ and $\Phi_{+}$ are
standard Bell states which are maximally $(\cA_{0},\, \cB_{0})$ -
entangled.
\subsection{Algebraic non - separability}
In the algebraic theory we can also simply characterize pure non -
separable  states. As it follows from Theorem \ref{sep}, this
property is shared by the states $\om$  whose restrictions
$\om_{\cA}$ and $ \om_{\cB}$ are non - pure. In this context,
maximal non - purity of $\om_{\cA}$ and $\om_{\cB}$, corresponds to
the maximal non - separability of the state $\om$.  The state
$\om_{\cA}$ is maximally non - pure if it is \textit{trace state}
i.e. if it satisfies
\begin{equation*}
\om_{\cA}(AA^{\prime})=\om_{\cA}(A^{\prime}A)
\end{equation*}
for all $A,\, A^{\prime}\in \cA$. When we apply this condition to
the generators of subalgebra $\cA$, we obtain
\begin{equation}
\om_{\cA}(A_{j}A_{k})=\om_{\cA}(A_{k}A_{j})=-\om_{\cA}(A_{j}A_{k}),\quad
j\neq k.\label{AjAk}
\end{equation}
Hence $ \om_{\cA}(A_{j}A_{k})=0$, and $ \om_{\cA}(A_{i})=0$ for
$i=1,2,3$. The same conclusion also follows for subalgebra $\cB$.
Thus  pure state $\om$ on $\Atot$ is maximally $(\cA,\, \cB)$ -
correlated  if
\begin{equation}
\om_{\cA}(A)=0\quad\text{and}\quad \om_{\cB}(B)=0\label{maxnonsep}
\end{equation}
for all $A\in \cA,\, A\neq \I$ and $B\in \cB,\, B\neq \I$.
\par
We can use the condition (\ref{maxnonsep}) to find maximally
entangled states for a given choice of subalgebras $\cA$ and $\cB$.
Let us give an example. Take the Bell pair $(\cA,\, \cB)$, defined
by (\ref{AB}) and consider the general pure state $\om_{\Psi}$. The
condition (\ref{maxnonsep}) applied to generators of $\cA$ and
$\cB$, gives the following equations for parameters $z_{i}$
\begin{equation}
\begin{split}
&\re\,\z{1}\,(\cz{3}-\cz{4})+\re\, \z{2}\,(\cz{3}+\cz{4})=0,\\
&\im\,\z{1}\,(\cz{4}-\cz{3})+\im\,\z{2}\,(\cz{3}+\cz{4})=0,\\
&|\z{2}|^{2}-|\z{1}|^{2}-2\re\,\z{3}\cz{4}=0.
\end{split}\label{zeroA},
\end{equation}
and
\begin{equation}
\begin{split}
&\re\, \z{1}\,(\cz{3}+\cz{4})+\re\,\z{2}\,(\cz{3}-\cz{4})=0,\\
&\im\,\cz{1}\,(\z{3}+\z{4})+\im\,\z{2}\,(\cz{3}-\cz{4})=0\\
&|\z{2}|^{2}-|\z{1}|^{2}+2\re\, \z{3}\cz{4}=0.
\end{split}\label{zeroB}
\end{equation}
We solve equations (\ref{zeroA}) and (\ref{zeroB}) and find that
\begin{equation*}
\begin{split}
&\z{1}=\frac{r}{\sqrt{2}}\,\cos\varphi,\\
&\z{2}=-\frac{r}{\sqrt{2}}\,\cos\varphi\:e^{2i\vartheta},\\
&\z{3}=r\sin\varphi\,e^{i\vartheta},\\
&\z{4}=i\,\sqrt{1-r^{2}}\:e^{i\vartheta},
\end{split}
\end{equation*}
where $r\in[0,1],\, \varphi\in [0,\pi/2],\, \vartheta\in [0,2\pi]$. So the maximally $(\cA,\, \cB)$ - correlated
states form a three - parameter family of vector states
\begin{equation}
\begin{split}
\Psi_{r,\varphi,\vartheta}=&\frac{r}{\sqrt{2}}\cos\varphi\,\eb{1}-\frac{r}{\sqrt{2}}
\cos\varphi\,e^{2i\vartheta}\,\eb{2}\\
&+\,r\sin\varphi\,e^{i\vartheta}\,\eb{3}+i\,\sqrt{1-r^{2}}\,e^{i\vartheta}\,\eb{4}.
\end{split}\label{ABmax}
\end{equation}
As we see, none of the standard Bell states belong to this family.
\subsection{Any pure state can be separable (or entangled)}
As we have shown in the previous sections, there are examples of two
- qubit states which are separable and at the same time entangled,
depending on the choice of Bell pair of subalgebras. Now
we show the general result: for any vector state $\om$, there exists
a Bell pair $(\cA,\, \cB)$ such that the corresponding total
correlation $C_{\om}(\cA,\,\cB)=c$, where $0\leq c \leq 1$ is a
given number. It means that the same state can be separable or
entangled to the given degree of non - separability. Although this
fact was already demonstrated (see e.g. \cite{HR}), our proof is
simpler and we are not using the  notion of tensor product.
\par
Let $\al$ be the $\ast$ - automorphism of $\Atot$ such that it does
not leave invariant subalgebras $\cA_{0}$ and $\cB_{0}$. Obviously
$(\al(\cA_{0}),\, \al(\cB_{0}))$ as well as
$(\al^{-1}(\cA_{0}),\al^{-1}(\cB_{0}))$ are the new Bell pairs. We
start with the simple observation: the total correlations in the
states $\om_{\al}=\om\circ\al$ and $\om$ are related by
\begin{equation}
C_{\om_{\al}}(\cA_{\al},\cB_{\al})=C_{\om}(\cA_{0},\,
\cB_{0}),\label{Cal}
\end{equation}
where
\begin{equation*}
\cA_{\al}=\al^{-1}(\cA_{0}),\quad
\cB_{\al}=\al^{-1}(\cB_{0}).\label{AalBal}
\end{equation*}
Let $\om=\om_{\Psi}$ be an arbitrary pure state and
let $\om_{0}$ be given by the vector $\Phi$ defined by equation
(\ref{Fi}). It was shown above that the state $\om_{0}$ is
$(\cA_{0},\, \cB_{0})$ - correlated, with the total correlation
equal to the given number $c$, where $ 0\leq c\leq 1$. As it is well known,  all pure states
of two qubits form the complex projective space $\C{\mathbb P}^{3}$, on which unitary matrices act
transitively (see e.g. \cite{GHL}). Let $U$ be
the unitary matrix such that $\Psi=U\Phi$ and define $\ast$ -
automorphism $\al$
\begin{equation*}
\al (A)=U^{-1}AU.
\end{equation*}
Then $\om=\om_{0}\circ \al$ and if we take $\cA=\cA_{\al}$,
$\cB=\cB_{\al}$, then by  (\ref{Cal})
\begin{equation*}
C_{\om}(\cA,\,\cB)=C_{\om_{0}}(\cA_{0},\,\cB_{0})=c.
\end{equation*}
So we have
\begin{thm}
Let $\om$ be an arbitrary pure state of two - qubits. We can always
choose a Bell pair  of subalgebras in such a way that the total
correlation in $\om$ is equal to a given number between $0$ and $1$.
\end{thm}
It is worth to stress that the Bell pair realizing  non - separability
of a fixed state is not unique. There are pure states which are separable  with respect to different choices of
subalgebras. There are also maximally entangled states for at least two different Bell pairs
(see examples discussed above).
\section{Conclusions}
We have studied entanglement properties of two qubits, using the
ideas of algebraic quantum mechanics. General theory of entanglement
is based on the properties of the algebra of physical observables
and its partitions representing the subsystems. The states, defined
as linear functionals on observables, are entangled if they give non
- vanishing correlations between independent subsystems. So the
entanglement is always relative to a particular set of physical
observables. We have shown that the universal measure of pure state
entanglement can be obtained by considering the total correlation
between subsystems in a given state. The value of this quantity can
be simply computed in terms of the norm of the corresponding
correlation matrix. Abstract algebraic characterization of
separability or non  - separability of any pure state can be
obtained by considering its restrictions  to subsystems. Purity (non
- purity) of restrictions are equivalent to separability (non -
separability) of a state.

\end{document}